\documentclass[12pt]{article}

\usepackage{graphicx,url}

\usepackage[utf8]{inputenc}  

\usepackage{authblk} 

\usepackage{amsmath,amsfonts,amssymb}
\usepackage{hyperref}
\usepackage{bm} 

\usepackage{pdfpages}

\title{Variable selection for (realistic) stochastic blockmodels}

\author[1]{Mirko Signorelli
\footnote{m.signorelli@lumc.nl\\
\textbf{Citation information.} Please cite as: Signorelli, M. (2017). 
\textit{Variable selection for (realistic) stochastic blockmodels}. In: Petrucci, A., Verde, R (editors), 
\textit{SIS 2017. Statistics and Data Science: new challenges, new generations}, pp. 927-934. 
Firenze University Press. ISBN: 978-88-6453-521-0.}}

\affil[1]{Department of Medical Statistics and Bioinformatics, Leiden University Medical Center (NL)}

\date{}

\begin{document} 

\maketitle

\vspace{-1cm}

\begin{abstract}
Stochastic blockmodels provide a convenient representation of relations between communities 
of nodes in a network. However, they imply a notion of stochastic equivalence that is often 
unrealistic for real networks, and they comprise large number of parameters that can make them 
hardly interpretable.
We discuss two extensions of stochastic blockmodels, and a recently proposed variable selection 
approach based on penalized inference, which allows to infer a sparse reduced graph summarizing 
relations between communities. 
We compare this approach with maximum likelihood estimation on two datasets on face-to-face 
interactions in a French primary school and on bill cosponsorships in the Italian Parliament.

\hspace{-0.1cm}

\end{abstract}

\section{Introduction}
\label{sec:1}
There is a long tradition in the study of graphs and relational data, whose origins can be arguably traced back to the seminal works of Moreno (1934) and Erd{\"o}s and R{\'e}nyi (1959). For decades, however, the study of real networks was limited by the difficulty to collect comprehensive data on large and complex systems. At the turn of the XX century, network science received a sudden boost from many technological advances that have facilitated the collection of relational data in a plentiful of fields. Examples include the advent of high throughput technologies in genetics and of functional magnetic resonance imaging in neuroscience, as well as the development of sensor-based measurements and the diffusion of social media in social network analysis.

The increasing availability of data on real networks has fostered research on their focal properties. These include the famous ``small-world property'', encapsulated in the idea of ``six degrees of separation'' between any two inhabitants of the Earth, and the idea that networks are scale free, i.e., that a few nodes in a network account for most of the connections therein. A further commonly observed feature of real networks is the presence of groups of nodes (``communities'') that are highly connected to each other, and poorly connected to the rest of the network. This \textit{community structure} may be induced by observed attributes of the nodes, or it could be thought as the result of an unknown latent factor. In this paper we focus on two extensions of stochastic blockmodels \textit{a priori}, a class of network models that allow to relate such community structures to observed attributes of the nodes\footnote{A related class of blockmodels is that of \textit{a posteriori} stochastic blockmodels (Wasserman and Anderson, 1987), whose aim is the detection of communities rather than the description of relations between known blocks of nodes.}. Although stochastic blockmodels are a convenient way to represent relations between groups of nodes in a network, they require a large number of parameters, which increases quadratically with the number of groups. As a consequence, when a large number of groups is considered, they typically yield cumbersome results that are hard to interpret. Signorelli and Wit (2017) proposed to address this issue by estimating stochastic blockmodels in a penalized likelihood setting. This allows to perform variable selection for stochastic blockmodels, to reduce model complexity and to derive a sparse reduced graph that summarizes the most important interactions within and between communities.

The paper is organized as follows. In Section \ref{sec:2} we discuss how the stochastic blockmodel can be extended so as to incorporate information on the degrees of nodes and on nodal or edge-specific covariates, and how to derive a reduced graph that summarizes relations between communities. Section \ref{sec:3} shortly introduces the variable selection approach proposed by Signorelli and Wit (2017). In Section \ref{sec:4}, we illustrate the proposed methodology with two examples on face-to-face contacts in a French high school, and on bill cosponsorship in the Italian Parliament.

\section{Representing community structure with realistic stochastic blockmodels}
\label{sec:2}
We consider an undirected graph $\mathcal{G} = (V, E)$, which features a set of edges $E \subseteq V \times V$ between a set of nodes or vertices $V = \{1,...,n\}$. We denote by $A$ the (symmetric) adjacency matrix of the graph, whose entries $a_{ij}$ are non-null if and only if an edge between nodes $i$ and $j$ is present, and we assume absence of self-loops, i.e., $a_{ii} = 0 \:\: \forall i \in V$. Moreover, we distinguish binary graphs, where $a_{ij} \in \{0,1\}$, from edge-valued graphs where $a_{ij} \in \mathbb{N}$, and we view each $a_{ij}$ as a draw from the random variable $Y_{ij}$.

\subsection{Stochastic blockmodel: definition and extensions}
A stochastic blockmodel assumes that a partition $\mathcal{P}$ of $V$ into $p$ blocks of nodes $\{B_1,...,B_p\}$ is available. According to the definition proposed by Holland et al. (1983), a network model is a \textit{stochastic blockmodel} if
\begin{itemize}
\item the random variables $Y_{ij}$ are independent;
\item $Y_{ij}$ and $Y_{kl}$ are identically distributed if nodes $i, k$ belong to the same block $B_r$, and nodes $j, l$ to the same block $B_s$.
\end{itemize}
This definition implies that every node within the same block is \textit{stochastically equivalent}, to wit, that it is possible to swap any two nodes that are members of the same block without affecting the probability distribution of the graph.

The assumption of stochastic equivalence within blocks represents a strong limitation of stochastic blockmodels. For example, it entails that the expected degree of nodes within a block is the same, whereas most real networks feature a strong heterogeneity in the distribution of node degrees. This was noted already by Wang and Wong (1987), who proposed to integrate the stochastic blockmodel with a set of nodal fixed effects. For undirected binary graphs, their \textit{degree-corrected blockmodel} assumes that if $i \in B_r$ and $j \in B_s$, then $Y_{ij} \sim Bern(\pi_{ij})$ and
\begin{equation}
\text{logit} \: \pi_{ij} = \beta_0 + \alpha_i + \alpha_j + \phi_{rs}, \label{wangwong-model}
\end{equation}
subject to the identifiability constraints $\sum _i \alpha_i = 0$ and $\sum _s \phi_{rs} = 0 \;\:\forall r \in \{1,...,p\}$. Here, a positive block-interaction effect $\phi_{rs}$ indicates that nodes in blocks $B_r$ and $B_s$ tend to interact preferentially with each other. Note that Equation \eqref{wangwong-model} breaks the assumption of stochastic equivalence of nodes within a block and, thus, the model proposed by Wang and Wong (1987) is not a stochastic blockmodel in the sense of Holland et al. (1983). However, it allows a more realistic description of a network with known block-structure: as a matter of fact, it takes into account both nodal information on the popularity or productivity of each node ($\alpha_i$ and $\alpha_j$), and information on the extent of interaction between pairs of blocks ($\phi_{rs}$).

A further limitation of stochastic blockmodels is that they postulate that the formation of edges depends only on block membership of the nodes. However, often it is reasonable to imagine that other factors besides block membership can affect the process of edge formation. Signorelli and Wit (2017) proposed an extension to stochastic blockmodels that allows the formation of an edge to depend both on block memberships, and on a set of nodal or edge-specific covariates $x_{ij}$. They considered the case of an undirected, edge-valued graph and viewed the formation of an edge between $i \in B_r$ and $j \in B_s$ as the result of a Poisson process whose rate depends both on blocks $B_r$ and $B_s$ and on the covariates $x_{ij}$. The resulting network model can be estimated with a generalized linear model where $Y_{ij} \sim Poi(\mu_{ij})$ and
\begin{equation}
\text{log} \: \mu_{ij} = \beta_0 + x_{ij} \beta + \gamma_r + \gamma_s + \phi_{rs}, \label{signowit-model}
\end{equation}
subject to the identifiability conditions $\sum _r \gamma_r = 0$ and $\sum _s \phi_{rs} = 0 \:\:\forall r \in \{1,...,p\}$. Likewise model \eqref{wangwong-model}, also model \eqref{signowit-model} breaks the assumption of stochastic equivalence within blocks. However, it differs from model \eqref{wangwong-model} in two aspects: it allows to account for factors other than group membership, and it replaces the nodal fixed effects $\alpha_i$ with block effects $\gamma_r$.

\subsection{How to derive a reduced graph}
The focal point of a stochastic blockmodel and of its (more realistic) extensions outlined above is their capacity to summarize a (potentially large) network by making some statements on the relations that exist between the blocks (Anderson, 1992). In particular, stochastic blockmodels make it possible to infer from the observed graph $\mathcal{G}$ a \textit{reduced graph} $\mathcal{G}_R = (\mathcal{P}, E_R)$ whose nodes are the blocks.

The reduced graph represents a synthetic way to visualize the relations that exist between blocks in the network. Typically, it is employed to show which blocks interact more with each other. For binary graphs, Anderson (1992) proposed to derive a reduced graph from a stochastic blockmodel by setting a threshold on the predicted interaction probability $\hat{\pi}_{rs}$ to observe an edge between nodes in blocks $B_r$ and $B_s$. However, the reduced graph obtained with this procedure arbitrarily depends on the choice of the threshold and, furthermore, it might display some blocks as connected to any other block, just because its nodes have, on average, high degrees. Moreover, this procedure does not directly generalize to the case of edge-valued graphs.

To overcome these problems, Signorelli and Wit (2017) derive the reduced graph in a different way, drawing an edge between two blocks $B_r$ and $B_s$ if the estimate $\hat{\phi}_{rs}$ of the corresponding block-interaction parameter $\phi_{rs}$ is positive. This approach is coherent with the parametrizations employed in models \eqref{wangwong-model} and \eqref{signowit-model}, where a positive $\phi_{rs}$ entails evidence of attraction between $B_r$ and $B_s$. Thus, the resulting reduced graph will display those pairs of blocks whose nodes tend to interact more with each other. The reduced graphs presented in Section \ref{sec:4} are obtained with this method.

\section{Variable selection for stochastic blockmodels}
\label{sec:3}
The description of relations between pairs of blocks provided by stochastic blockmodels requires the use of a rather large number of parameters. This is necessary in order to model each interaction between blocks $(B_r, B_s)$, $\:s \geq r \in \{1,...,p\}$. In particular, model \eqref{wangwong-model} includes $q_1 = n + p(p-1)/2$ parameters, and model \eqref{signowit-model} $q_2 = \text{dim}(\beta) + p(p+1)/2$. As we will show in Section \ref{sec:4}, when many blocks are considered ($p \geq 10$) this often yields reduced graphs with a plentiful of links that are cumbersome to interpret.

In a study on collaborations between Italian political parties, Signorelli and Wit (2017) analysed bill cosponsorship networks in the Chamber of Deputies with model \eqref{signowit-model} and observed that although positive and negative estimates $\hat{\phi}_{rs}$ respectively entail collaboration and repulsion between Deputies in parties $B_r$ and $B_s$, it is also possible to imagine a situation of indifference between collaboration and repulsion for some pairs of parties. This indifference directly corresponds to $\phi_{rs} = 0$ in models \eqref{wangwong-model} and \eqref{signowit-model}. However, with maximum likelihood estimation it is highly unlikely that any of the point estimates $\hat{\phi}_{rs}$ will be exactly zero. For this reason, they advocated the penalization of the block-interaction terms $\phi_{rs}$ (as well as of the covariate vector $\beta$ in Equation \eqref{signowit-model}) and employed the adaptive lasso (Zou, 2006) to estimate their model.

This penalized inference approach yields two advantages: on the one hand, it allows to distinguish situations of indifference between blocks from collaborations or repulsions; on the other hand, it is capable to reduce the complexity of the inferred model by shrinking some of its parameters to 0. As a result, it enables to infer a sparse reduced graph, which is typically easier to interpret than the one based on maximum likelihood estimation.

We remark that because of the identifiability conditions that ought to be imposed in models \eqref{wangwong-model} and \eqref{signowit-model}, $p$ block-interaction parameters do not directly appear in the models and, thus, they cannot be penalized. Given that the parameters for interactions within each block, $\phi_{rr}$, are anyway likely to be positive, we substitute $\phi_{rr} = - \sum _{s \neq r} \phi_{rs}$ for every $r\in \{1,...,p\}$ in \eqref{wangwong-model} and \eqref{signowit-model}. By doing so, we penalize each block-interaction parameter $\phi_{rs}$ ($r \neq s$), and derive each $\phi_{rr}$ from the constraints.

In the next Section we consider two examples of penalized inference for stochastic blockmodels, and carry out a comparison of this approach with the one based on maximum likelihood.

\section{Applications}
\label{sec:4}

\subsection{Face-to-face contacts in a French primary school}
We consider data on face-to-face interactions in a French primary school collected by Stehl{\'e} et al. (2011). The study, which lasted 2 days, employed sensors to detect face-to-face interactions between students and teachers that lasted at least 20 seconds. Here, we focus on the interactions measured in the first day and consider a binary graph whose nodes are students and teachers, and where an edge between two nodes indicates that at least an interaction between them was recorded during the day.

The school comprises 10 classes (2 for each level). The available information for each node is its status (student or teacher); furthermore, for students also class and gender are known. Thus, we partition the nodes into 21 blocks: 20 blocks partition students according to their class and gender, and the last one contains teachers.

We employ model \eqref{wangwong-model} to study the pattern of interactions among the blocks, and compare the reduced graphs that can be derived by employing maximum likelihood, and the penalized likelihood estimation procedure described in Section \ref{sec:3}.

\begin{figure}[t]
\centering
\includegraphics[scale=0.37, trim = {3cm 1cm 2cm 1cm}]{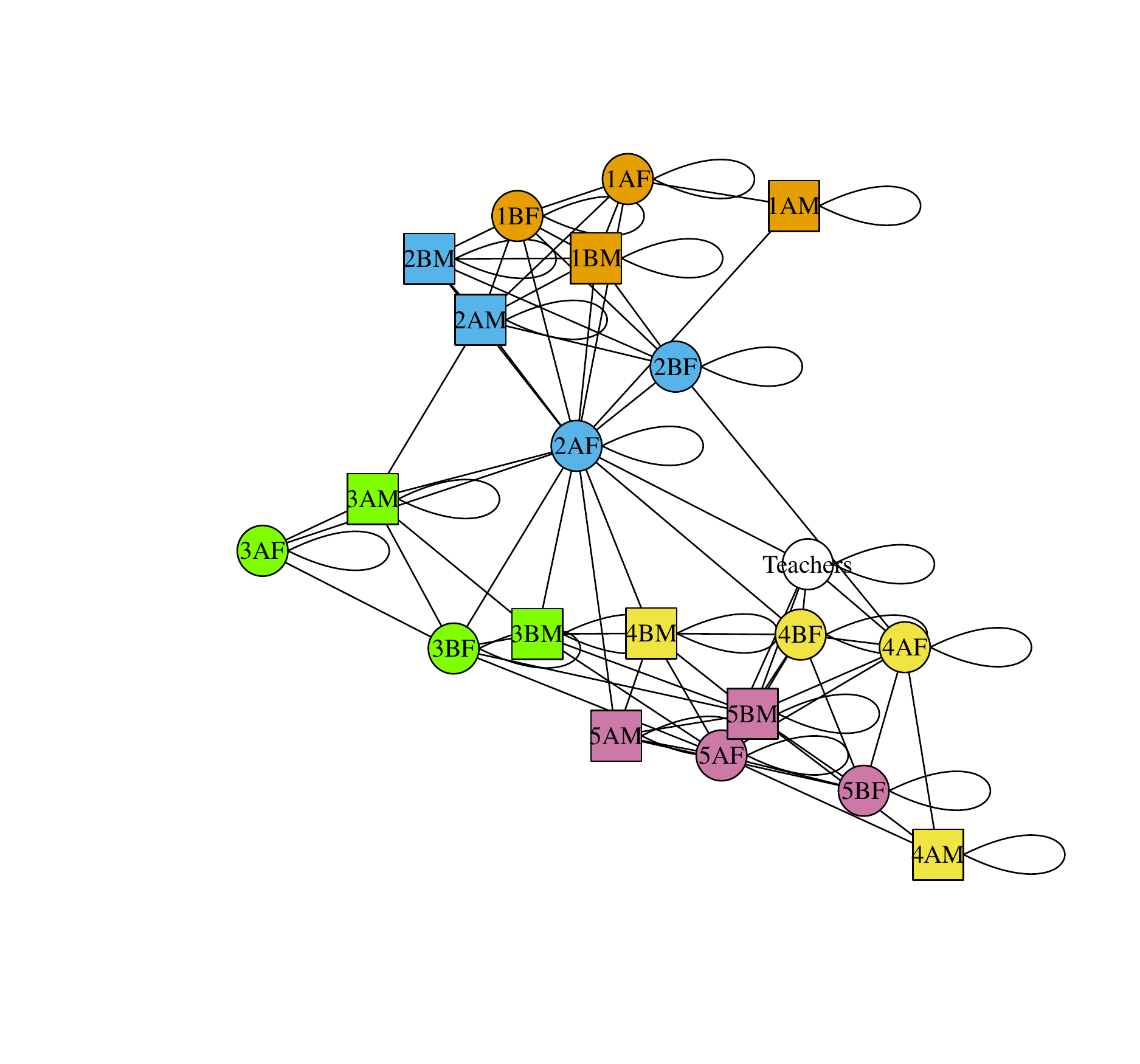}
\includegraphics[scale=0.37, trim = {2cm 1cm 3cm 1cm}]{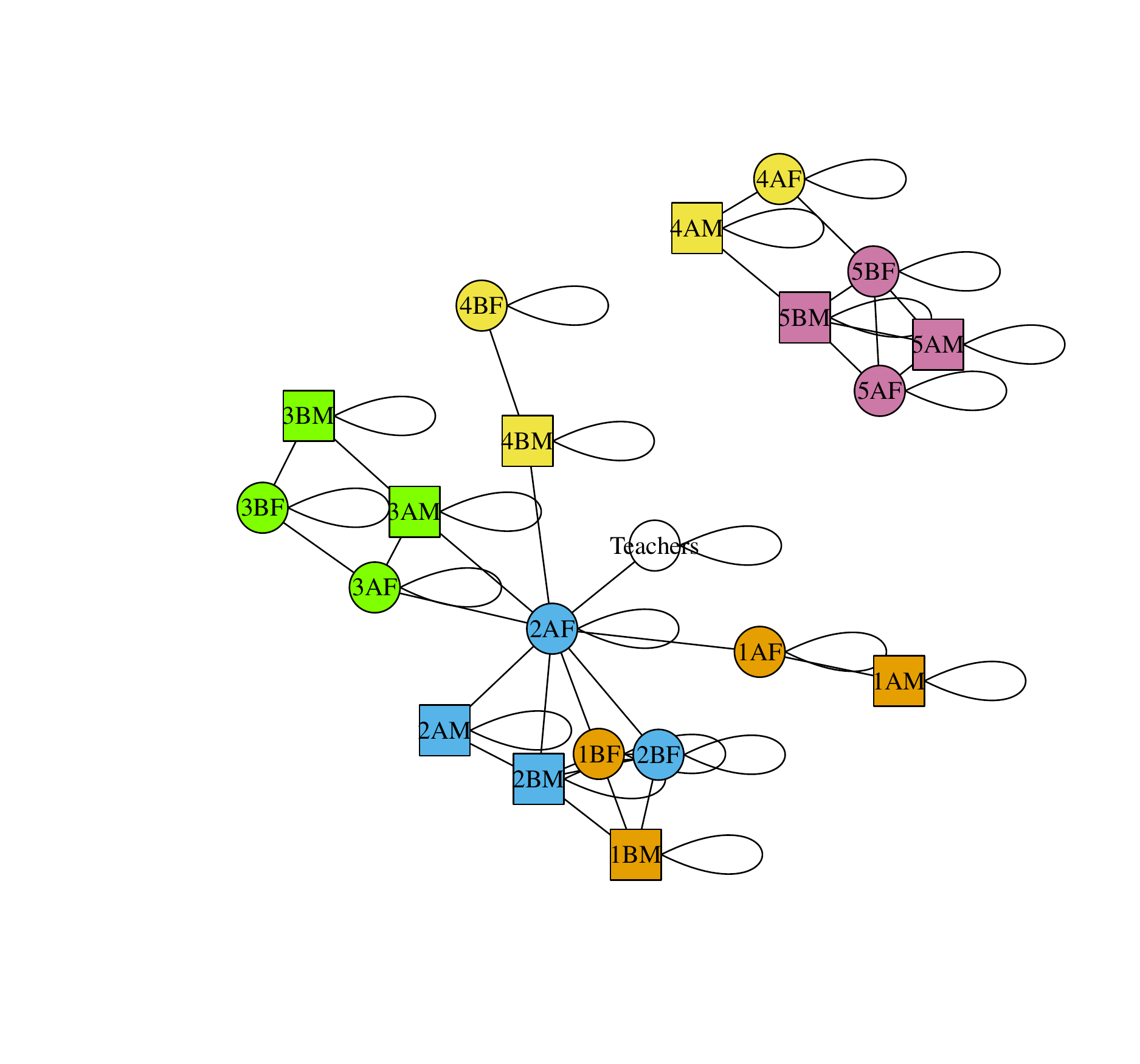}
\put (-266,175) {{\footnotesize{Reduced graph based on the}}}
\put (-270,165) {{\footnotesize{maximum likelihood estimator}}}
\put (-107,175) {{\footnotesize{Reduced graph based on the}}}
\put (-100,165) {{\footnotesize{adaptive lasso estimator}}}
\caption{Comparison of reduced graphs based on maximum likelihood and 
penalized likelihood inference, displaying interactions between groups 
of students (and teachers) in a French primary school. Node colors denote
 grades and their shapes distinguish female (circle) from male (square) 
 students. The label of each block indicates the grade (1-5), the section
  (A or B) and the gender (F or M) of students. The white circular node 
  indicates the block of teachers.}
\label{fig1}
\end{figure}

Maximum likelihood estimation results into 86 positive, and 145 negative, estimates of the block-interaction parameters. As a result, the reduced graph in Figure \ref{fig1} displays a large number of interactions between the blocks. Penalized likelihood estimation, instead, shrinks 88 block-interaction parameters to 0, resulting into 52 positive and 91 negative parameter estimate $\hat{\phi}_{rs}$. A direct consequence of this is that the reduced graph displaying interactions between blocks is now more readable. In particular, the presence of self-loops indicates that members within each block interact frequently with their peers. Furthermore, a link is present between male and female students within each class. Whereas students in their fifth grade also interact across classes in their same grade (5A and 5B) irrespective of gender, the pattern of interaction between the two third grade classes (3A and 3B) seems to be affected by gender identity (males in 3A interact with males in 3B, and females in 3A with females in 3B). Instead, we do not find any interaction between first, or fourth grade classes (1A-1B and 4A-4B, respectively).

\subsection{Bill cosponsorship in the Italian Parliament}
Signorelli and Wit (2017) employed data on bill cosponsorships to reconstruct the pattern of collaborations between Italian political parties in the Chamber of Deputies from 2001 to 2015. Here we focus our attention on the bill cosponsorship network for the first part of the XVII legislature (2013 - 2015) and make a comparison between maximum likelihood and penalized likelihood inference.

\begin{table}
\label{tab:1}
\caption{Comparison of maximum likelihood and adaptive lasso estimators for the parameter vector $\beta$ in model \eqref{signowit-model}. The reference modes are interactions between two male deputies (male-male) for gender effects, and between two junior deputies (junior-junior) for seniority.}
	\vspace{0.4cm}	\centering
		\begin{tabular}{l|c|c}
Covariate & Maximum likelihood & Adaptive lasso  \\
 & estimate & estimate\\ \hline
Intercept ($\beta_0$) & -3.83 & -3.86\\
Female-male interaction & 0.233 & 0.210\\
Female-female interaction & 0.659 & 0.634\\
Same electoral constituency & 0.550 & 0.554\\
Age difference & -0.011 & 0\\
Junior-senior interaction & 0.253 & 0.234\\
Senior-senior interaction & 0.700 & 0.712\\ 
		\end{tabular}
\end{table}

We define a bill cosponsorship network where a weighted undirected edge is present between two deputies if they have cosponsored together at least one bill. Edge weights represent the number of bills that each pair of deputies has cosponsored. During the XVII legislature, 10 parliamentary groups are represented in the Chamber. Those groups form the blocks in model \eqref{signowit-model}, where we furthermore consider covariates for gender, age and seniority of deputies, besides a dummy variable that indicates whether two deputies have been elected in the same electoral constituency. In the penalized model, we penalize each of the covariates and the block-interaction terms, and we employ the adaptive lasso for estimation.

Table \ref{tab:1} compares the results for the intercept $\beta_0$ and the parameter vector $\beta$. Here, the only (slight) difference is that the parameter for age difference is shrunk to 0 with the adaptive lasso. The other variables indicate that female and senior deputies are more active in cosponsorships, and that geographic proximity also increases the tendency to collaborate.

\begin{figure}
\centering
\includegraphics[scale=0.37, trim = {3cm 1cm 2cm 1cm},page = 1]{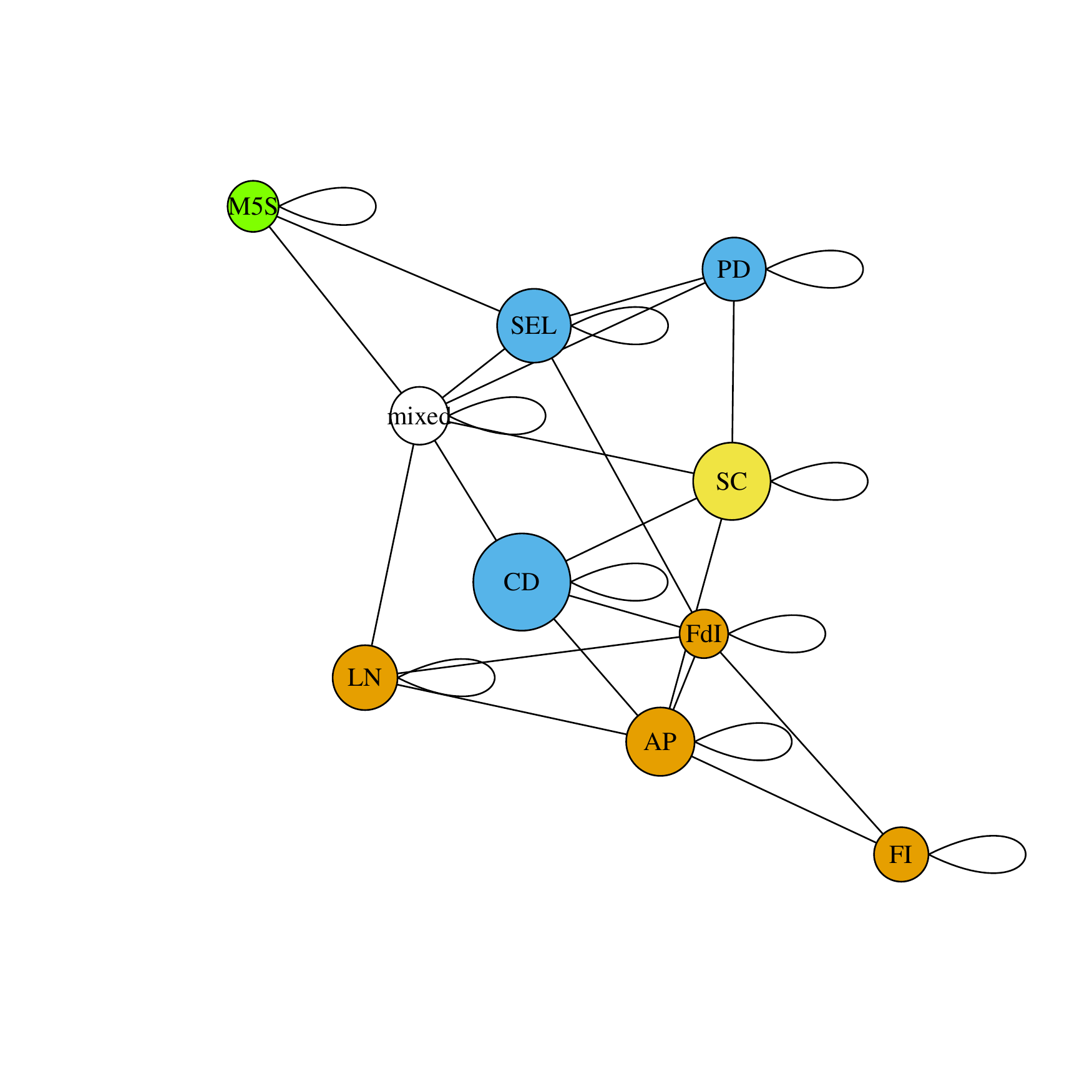}
\includegraphics[scale=0.37, trim = {2cm 1cm 3cm 1cm}, page = 2]{chamber-redgraphs.pdf}
\put (-266,175) {{\footnotesize{Reduced graph based on the}}}
\put (-270,165) {{\footnotesize{maximum likelihood estimator}}}
\put (-107,175) {{\footnotesize{Reduced graph based on the}}}
\put (-100,165) {{\footnotesize{adaptive lasso estimator}}}
\caption{Comparison of reduced graphs based on maximum likelihood and 
penalized likelihood inference, displaying collaborations between Italian 
political parties. Node size is proportional to group productivity. The 
colour of nodes is lightblue for left-wing parties, orange for right-wing 
ones, yellow for ``Scelta Civica'', green for ``Movimento 5 Stelle'' and 
white for the mixed group.}
\label{fig1}
\end{figure}

The main difference between the two approaches lies in the estimation of the block-interaction parameters $\phi_{rs}$. Maximum likelihood yields 29 positive, and 26 negative, estimates of the block-interaction parameters; the adaptive lasso, instead, shrinks 16 of those parameters to 0, resulting into 21 positive, 16 null and 18 negative estimates. Once more, the reduced graph of collaborations based on maximum likelihood is rather cumbersome to interpret, whereas the one based on the adaptive lasso is more readable. In particular, the latter points out collaborations within each party, between the 4 right-wing parties (orange), between three parties (`Centro Democratico', `Scelta Civica' and `Area Popolare') that belong to different coalitions, between the two main left-wing parties, and that deputies in the `mixed group' tend to collaborate with left-wing parties and with the `Movimento 5 Stelle'.

\section*{References}
\footnotesize
\noindent Anderson, C. J., Wasserman, S., Faust, K.: Building stochastic blockmodels. Soc. Netw., \textbf{14}(1), 137–-161 (1992).\vspace{0.15cm}

Erd{\"o}s, P., R{\'e}nyi, A.  On random graphs I. Publ. Math. (Debr.), \textbf{6}, 290-–297 (1959).\vspace{0.15cm}

Holland, P. W., Laskey, K. B., Leinhardt, S.: Stochastic blockmodels:
First steps. Soc. Netw., \textbf{5}(2), 109-–137 (1983).\vspace{0.15cm}

Moreno, J. L.: Who shall survive? Nervous and mental disease monograph series, \textbf{58} (1934).\vspace{0.15cm}

Signorelli, M., Wit, E. C.: A penalized inference approach to stochastic
block modelling of community structure in the Italian Parliament. J. Royal Stat. Soc., Ser. C (2017).\vspace{0.15cm}

Stehl{\'e}, J., Voirin, N., Barrat, A., Cattuto, C., et al.: High-resolution measurements of face-to-face contact patterns in a primary school. PloS one, \textbf{6}(8), e23176 (2011).\vspace{0.15cm}

Wang, Y. J., Wong, G. Y.: Stochastic blockmodels for directed graphs.
J. Am. Stat. Assoc., \textbf{82}(397), 8–-19 (1987).\vspace{0.15cm}

Wasserman, S., Anderson, C.: Stochastic a posteriori blockmodels: construction and assessment. Soc. Netw., \textbf{9}, 1--36.\vspace{0.15cm}

Zou, H.: The adaptive lasso and its oracle properties. J. Am.
Stat. Assoc., \textbf{101}(476), 1418-–1429 (2006).

\end{document}